\documentclass[prc,aps,twocolumn,preprintnumbers,amsmath,amssymb,floatfix,superscriptaddress]{revtex4}
\usepackage[usenames]{color}
\usepackage{amsmath}
\usepackage{graphicx}
\usepackage{float,epsfig}
\usepackage{lscape}
\newcommand\be{\begin{equation}}
\newcommand\ee{\end{equation}}
\newcommand\bea{\begin{eqnarray}}
\newcommand\eea{\end{eqnarray}}

\usepackage[usenames]{color}
\usepackage{ulem}

\begin{document}
\title{A Lee-Yang--inspired functional with a density--dependent neutron-neutron scattering length}

\author{M. Grasso}
\affiliation{IPNO, CNRS/IN2P3, Universit\'e Paris-Sud, Universit\'e Paris-Saclay, F-91406, Orsay, France}
\author{D. Lacroix}
\affiliation{IPNO, CNRS/IN2P3, Universit\'e Paris-Sud, Universit\'e Paris-Saclay, F-91406, Orsay, France}
\author{C.J. Yang}
\affiliation{IPNO, CNRS/IN2P3, Universit\'e Paris-Sud, Universit\'e Paris-Saclay, F-91406, Orsay, France}
\begin{abstract}
Inspired by the low--density Lee-Yang expansion for the energy of a dilute Fermi gas of density $\rho$ and momentum $k_F$, we introduce here a Skyrme--type functional that contains only $s$-wave terms and provides, at the mean--field level, (i) a satisfactory equation of state for neutron matter from extremely low densities up to densities close to the equilibrium point, and (ii) a good--quality equation of state for symmetric matter at density scales around the saturation point. This is achieved by using a density--dependent neutron-neutron scattering length $a(\rho$) which satisfies the low--density limit (for Fermi momenta going to zero) and has a density dependence tuned in such a way that the low--density constraint $|a(\rho) k_F| \le 1$ is satisfied at all density scales.  
\end{abstract}

\maketitle

The interaction between the constituents of very dilute Fermi systems is accurately determined by a few parameters associated to $s$--wave scattering processes. An example is given by ultracold trapped Fermi gases where the interaction may be reasonably well approximated by a zero--range force with a coupling constant directly related to the $s$--wave scattering length $a$ \cite{brunn,stringari,Blo08,Zwe11}. 
The study of a dilute regime in fermionic systems is of particular interest for a wide community of many--body practitioners for instance in the domains of atomic (cold fermionic trapped atoms) and nuclear (nuclear matter) physics, as well as in nuclear astrophysics for the investigation of the properties of neutron star crusts. Within such a wide framework, bridging Effective Field Theories (EFTs), which by construction correctly describe low--density regimes with energy--density--functional (EDF) theories, which are currently employed for instance in the nuclear many--body problem, is a very appealing challenge which requires a tight interchange and connections between EFT and EDF expertise and competences.
The dilute regime is characterized by the relation $|ak_F| < 1$.   
For such a regime,  Lee and Yang introduced in the 50s  
an expansion in $(ak_F)$ for the ground--state energy \cite{lee}. 
It is important to notice that, at the unitarity limit, for example in ultracold atomic dilute gases close to Feshbach resonances, another expansion is used, on $1/(ak_F)$ (instead of $ak_F$).
The first terms of the Lee and Yang low--density expansion in $(ak_F)$ are reported for instance in Refs. \cite{efimov1,efimov2,baker,bishop}. More recently, such terms were derived in the framework of EFTs \cite{hammer}. The first four terms contain only $s$--wave parameters, the $s$--wave scattering length and the associated effective range $r_s$ (the following term appearing in the expansion contains the $p$--wave scattering length). We report here the first four terms, in the case where the spin degeneracy is equal to 2 (for example for neutron matter),  
\begin{widetext}
\begin{center}
\begin{eqnarray}
\frac{E}{N}&=&\frac{\hbar^2k_F^2}{2m} \left[ \frac{3}{5}+ \frac{2}{3\pi}(k_Fa)+
 \frac{4}{35\pi^2}(11-2 \ln 2)(k_Fa)^2  + \frac{1}{10\pi} (k_F r_s) (k_F a)^2 + 0.019  (k_F a)^3 \right],
\label{ly}
\end{eqnarray}
\end{center}
\end{widetext}
where $N$ is the number of particles. Within EFT, the above equation is obtained via dimensional regularization (DR) with minimal subtraction. It is independent of the adopted regularization scheme, provided that a matching to an effective--range expansion is performed \cite{fu99}. 

The comparison of the Lee-Yang (LY) energy, Eq. (\ref{ly}), with the energy obtained in a many--body perturbative expansion indicates that 
the terms in $(k_F a)$, $(k_F a)^2$, and $(k_F a)^3$ correspond respectively to the leading--, second--, and third--order contributions produced by a zero--range interaction with a coupling constant related to the scattering length $a$ \cite{yang2016}. The term in $(k_F r_s) (k_F a)^2$ corresponds to the leading--order contribution provided by a velocity--dependent zero--range $s$-wave interaction. Furthermore, it was shown in Ref. \cite{yglo} that the $(k_F a)^2$--term may be alternatively obtained at leading order with a specific density--dependent zero--range force. We notice finally that the term in $(k_F a)^3$ has the same $k_F$--dependence as the term in $(k_F r_s) (k_F a)^2$. This implies that 
a zero--range $s$--wave velocity--dependent term
 may mimic such a term at leading order.   
At very small densities, these first terms of the LY expansion, containing only $s$--wave scattering parameters, are enough to correctly describe the energy of the system. 

Inspired by this expansion, we introduce here a Skyrme--type functional \cite{sky1,sky2}, containing only $s$-wave terms, that leads, at the mean--field level, to an EOS for neutron matter given by Eq. (\ref{ly}), with the relations 
\begin{eqnarray}
\nonumber
t_0(1-x_0)&=&\frac{4\pi\hbar^2}{m}a, \\
t_3(1-x_3)&=&\frac{144 \hbar^2}{35 m} (3\pi^2)^{1/3} (11-2 \ln 2)a^2, \\
\nonumber
t_1 (1-x_1) &=& \frac{2\pi\hbar^2}{m}(a^2 r_s + 0.19 \pi a^3),
\label{para}
\end{eqnarray}
where $t_0, t_1, t_3$, and $x_0, x_1, x_3$ are Skyrme parameters and the power of the density--dependent $t_3$-term is chosen equal to 1/3 \cite{yglo}. We require that 
such a functional: (i) correctly describes neutron matter from extremely low densities up to densities close to the equilibrium point of symmetric matter, $\rho=0.16$ fm$^{-3}$; (ii) provides in addition a correct equation of state (EOS) for symmetric matter at density scales around the saturation point. The requirement (i) cannot be fully satisfied by using the value of -18.9 fm for the $s$-wave scattering length. Such a very large value leads indeed to a correct description of neutron matter with the first terms of Eq. (\ref{ly}) only at extremely low densities \cite{yglo}. 
For this reason, resumed expressions have been proposed in the literature within EFT for cases where the scattering length is very large (see for instance Refs. \cite{steel,schae,kai}). Recently, going towards this direction, a hybrid functional was introduced, YGLO, combining a resumed expression (that guarantees the correct low--density behavior) and good properties of Skyrme--type forces, 
which are known to well describe the EOS of matter close to the equilibrium point of symmetric matter \cite{yglo}. A resumed  
functional making connection between cold atoms and neutron matter was introduced in Ref. \cite{Lac16}.  

In this work, we probe the possibility of adopting a simpler functional, which does not contain any resumed expression. To satisfy both requirements (i) and (ii) we impose that the neutron-neutron scattering length is density dependent, $a(\rho)$, in such a way to ensure the correct low--density behavior for neutron matter (for $k_F$ going to zero) and to justify the use of a LY--type EOS truncated at the very first terms (only parameters related to the $s$-wave scattering length are taken into account). 

Owing to the fact that the parameters $x_i$ associated to $s$--wave terms do not appear in the mean--field EOS of symmetric matter with a Skyrme force, we adjust here the parameters $t_i$ to have a reasonable mean--field EOS for symmetric matter, and we tune the neutron-neutron scattering length in the following way: We impose that it takes the value of -18.9 fm up to a Fermi momentum $k_F^{max}$ such that $k_F^{max} 18.9 =1$. This ensures the correct low--density behavior. It turns out that $k_F^{max} \sim 0.05$ fm$^{-1}$, corresponding to a maximum density  $\sim 4 \times 10^{-6}$ fm$^{-3}$, where the density and the Fermi momentum are related by the relation $k_F=(3\pi^2 \rho)^{1/3}$. Beyond this density value, 
we generalize the low--density constraint $|k_F \; a|\leq 1$ (that identifies the density window where the LY formula is valid) to the case where the scattering length is density dependent, with the 
relation $|k_F \; a(\rho)|\leq 1$ used to tune the density dependence of the scattering length. 
 Interestingly, the momentum dependence $1/k_F$ obtained by imposing such a constraint strongly resembles to the magnetic--field dependence of the scattering length 
in ultracold trapped atoms close to Feshbach resonances \cite{stringari}. 
This indicates a very strong analogy with ultra--cold atomic gases, where the scattering length is tuned by an applied magnetic field. Our strategy of using a Fermi momentum (or density)--tuned scattering length leads in our case to a strikingly similar behavior.

We plot in Fig. 1 the neutron-neutron scattering length as a function of the density (a) and as a function of the Fermi momentum (b). The lower curve corresponds to the tuning $|k_F \; a(k_F)|=1$. As an illustration of the sensitivity to the hypothesis discussed above, the upper curve delimiting the green region is also shown, corresponding to a tuning obtained by imposing a stricter low--density constraint, $|k_F \; a(k_F)|=0.5$. The green areas in the two panels of the figure contain all the intermediate cases.  When replacing $a$ by $a(\rho)$, 
the  $x_i$ parameters become density dependent and their expressions may be deduced from Eqs. (2).

\begin{figure}
\includegraphics[scale=0.35]{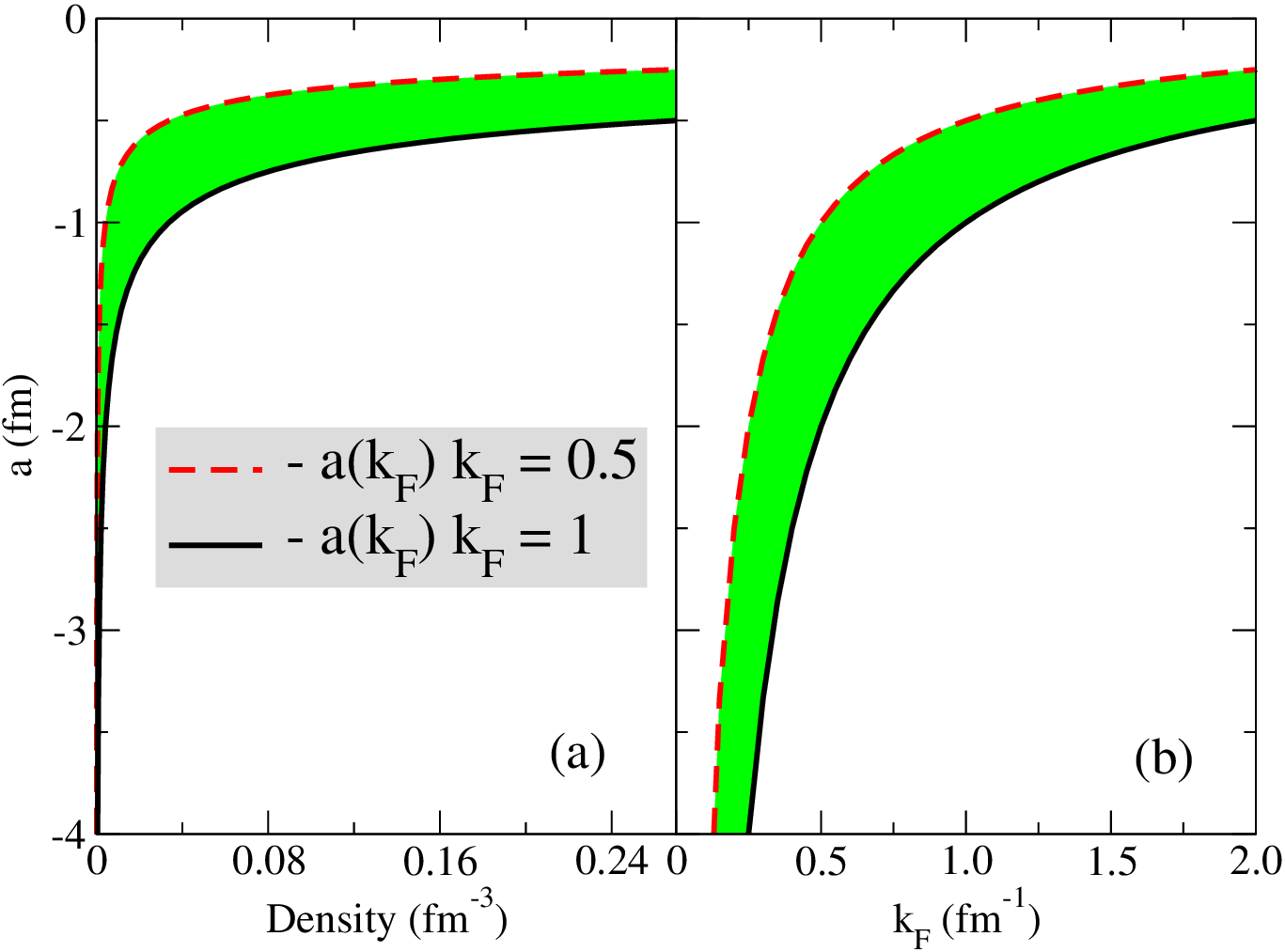}
\caption{Neutron-neutron $s$--wave scattering length as a function of the density (a) and of the Fermi momentum (b). The green region 
contains all the possible cases between $|k_F \; a(k_F)|=0.5$ (red dashed line) and $|k_F \; a(k_F)|=1$ (black solid line).}
\label{arhok}
\end{figure}


{\bf{First two terms of the Lee-Yang expansion.}} 

The EOS obtained by dropping the last two terms of Eq. (\ref{ly}) corresponds to a mean--field EOS within a Skyrme $t_0-t_3$ model. 
Within such a simplified model, we adjusted in Ref. \cite{yglo} the parameters $t_0$ and $t_3$ to have a satisfactory EOS for symmetric matter around the saturation point, providing a saturation density of 0.16 fm$^{-3}$ with an energy per particle of -16.04 MeV: $t_0$ ($t_3$) = -1803.93 MeV fm$^3$ (= 12911.00 MeV fm$^4$). 
By using these values for $t_0$ and $t_3$, we may deduce the density dependence of the parameters $x_0$ and $x_3$ through the density dependence of the neutron-neutron scattering length (Fig. 2). 
Notice that the values of the parameters $x_0$ and $x_3$ vary from 
-4.46 and -139.40, respectively, at zero density \cite{yglo}, to  values which are closer to typical $x_0$, $x_3$ values in Skyrme forces at larger densities. 
\begin{figure}
\includegraphics[scale=0.35]{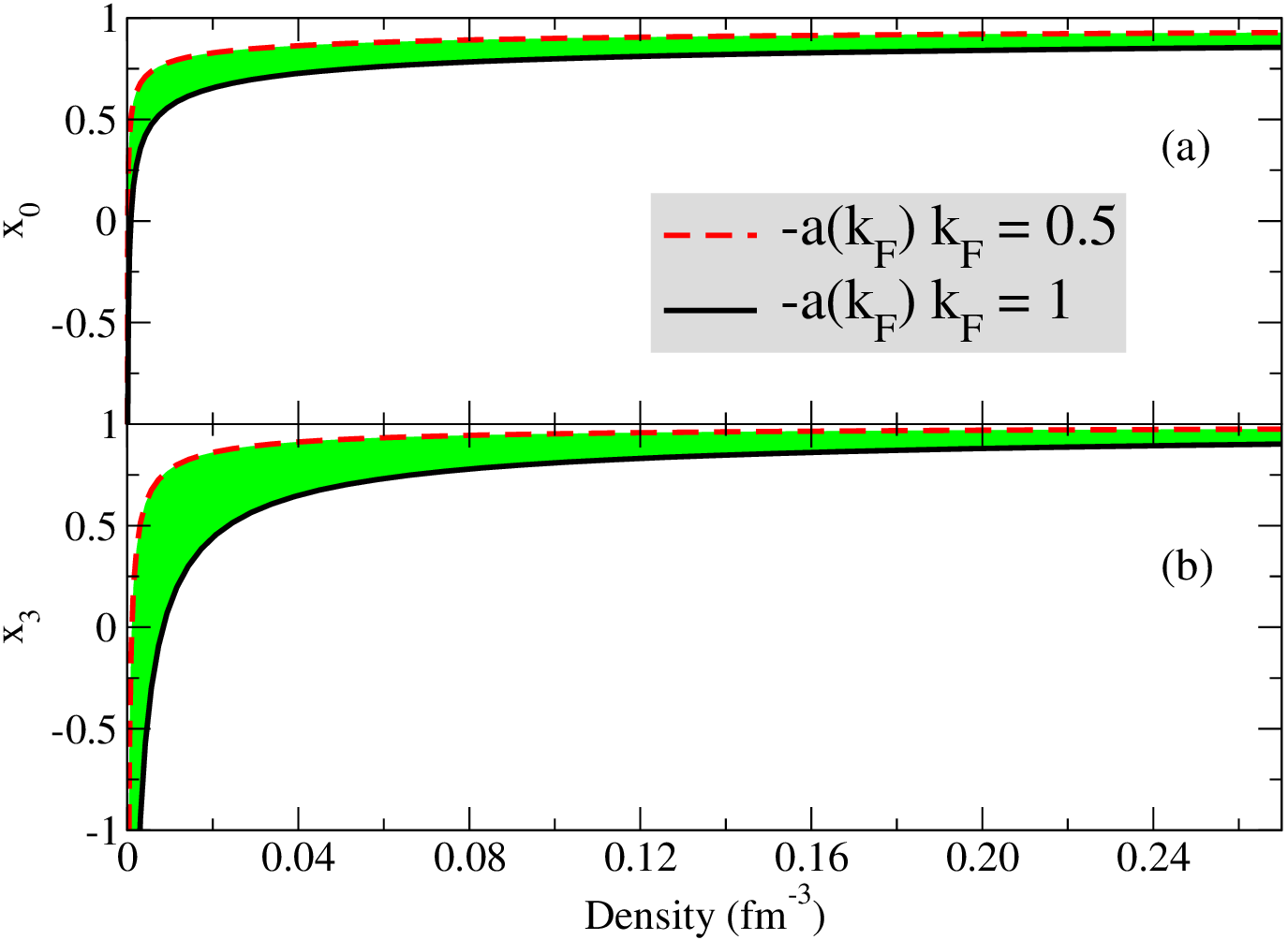}
\caption{Density dependence of the parameters $x_0$ (a) and $x_3$ (b). The green region 
contains all the possible cases between $|k_F \; a(k_F)|=0.5$ (red dashed line) and $|k_F \; a(k_F)|=1$ (black solid line).}
\label{x0x3}
\end{figure}
The corresponding EOS of symmetric matter is plotted in Fig. 3 and compared to the SLy5--mean--field \cite{sly5} EOS (used as a benchmark for the fit in Ref. \cite{yglo}). 
\begin{figure}
\includegraphics[scale=0.35]{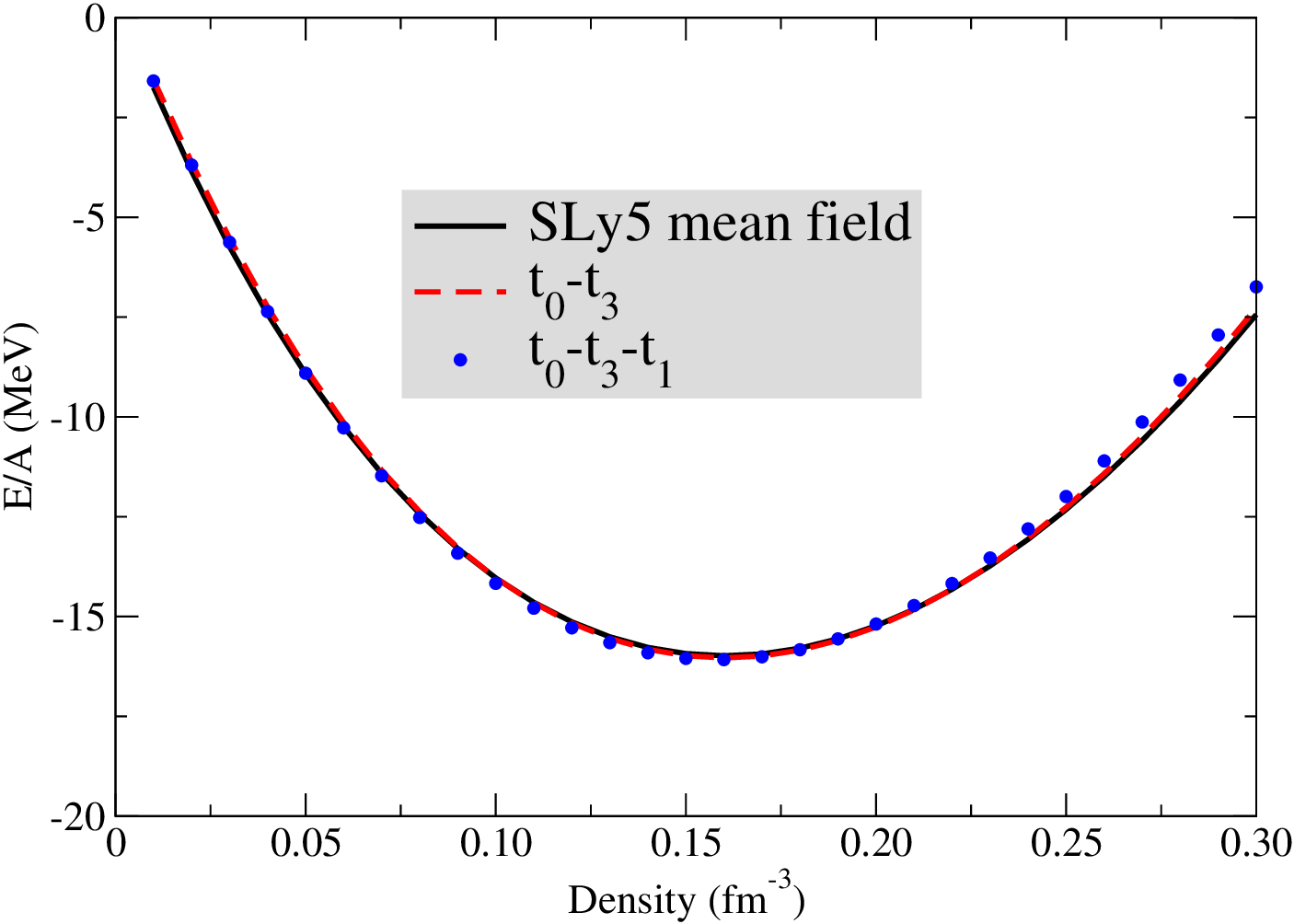}
\caption{EOS of symmetric matter obtained within a $t_0-t_3$ model, with the parameters adjusted in Ref. \cite{yglo} (red dashed line) and within a $t_0-t_3-t_1$ model, with the parameters adjusted in this work (blue circles). For comparison, the SLy5-mean-field EOS is plotted (black solid line).}
\label{eossym}
\end{figure}

\begin{figure}
\includegraphics[scale=0.35]{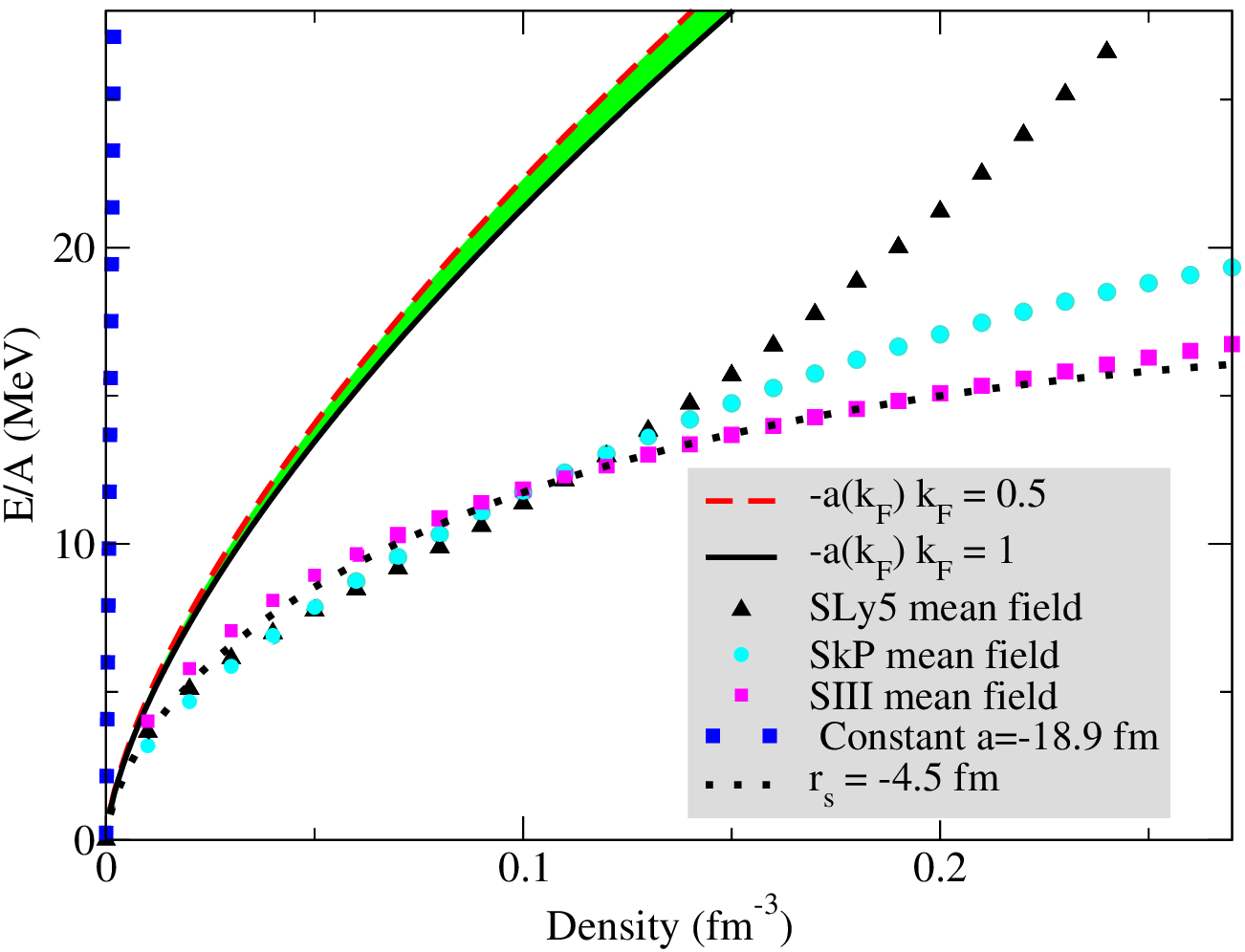}
\caption{EOS of neutron matter as a function of the density computed in the mean--field approximation with SLy5 (black triangels), SkP (cyan circles), and SIII (magenta squares). The EOSs obtained with the first two terms of the Lee-Yang expansion by using the constant value $a=-18.9$ fm (blue squares) and a density--dependent scattering length are also shown. For the latter case, the black solid line represents the EOS obtained with the low--density constraint  $|k_F \; a(k_F)|=1$ whereas 
the red dashed curve represents the EOS obtained by imposing $|k_F \; a(k_F)|=0.5$. The green area contains the intermediate cases. The black dotted line represents the EOS obtained by imposing $|k_F \; a(k_F)|=1$ and using an effective range of -4.5 fm.}
\label{eosrho}
\end{figure}

The EOS for neutron matter with $x_0$, $x_3$ displayed in Fig. \ref{x0x3} is shown in Fig. \ref{eosrho}, where also the 
SLy5--mean--field EOS is drawn for comparison (black triangles). 
We also show in the same figure two alternative mean--field Skyrme EOSs, obtained with the parameterizations SkP \cite{skp} (cyan circles) and SIII \cite{siii} (magenta squares). Whereas the SLy5 parametrization was designed to accurately 
reproduce a microscopic EOS for neutron matter even beyond the saturation point of symmetric matter, the other two Skyrme parameterizations were not adjusted in the same way and are not so accurate. However, they still provide reasonable results for neutron matter at least up to densities around the equilibrium point of symmetric matter. 
One observes that the upper and lower EOSs delimiting the green area differ very weakly. 
To estimate how much the density dependence of the scattering length affects the EOS, we also plot the EOS obtained by using a constant scattering length equal to the free value, -18.9 fm, (blue squares) which obviously leads to a totally wrong curve, except at extremely small densities. 

One observes that the obtained EOS is still quite far from the Skyrme EOSs.  The energy is systematically too high  indicating 
that an attractive contribution is missing.  

{\bf{Including the $s$--wave $k_F^5$ terms. }}

The value of the effective range associated to the scattering length $a=$-18.9 fm is 2.75 fm. In general, the term containing the effective range may be neglected in the LY expansion if $k_F |r_s| < 1$. For higher momenta,   
$k_F |r_s| \gtrsim 1$ and the corresponding term cannot be neglected anymore. In our case, the scattering length is equal to -18.9 fm only up to $k_F \sim $ 0.05 fm$^{-1}$. At this value of the Fermi momentum and for $r_s=$ 2.75 fm, $k_F r_s \sim $ 0.14, which is still sensibly less than 1. Thus, up to  $k_F \sim $ 0.05 fm$^{-1}$, the effective--range term may be safely neglected. However, at higher densities, the value of the scattering length changes very fast as a function of the density and it would be meaningless to still associate to such a value an effective range of 2.75 fm. 
Furthermore,   
we have observed in the Skyrme $t_0-t_3$ model that, at ordinary nuclear densities, a density--dependent neutron-neutron scattering length is not sufficient to reproduce a reasonable EOS and that an attractive contribution is missing. 
Such a  missing attractive contribution in the EOS may indeed be obtained by including the following two terms of the expansion, which are still $s$--wave terms and have both a $k_F^5$ dependence (Eq. (2) shows the relation with the $t_1$ velocity--dependent term of a Skyrme model). The first of these terms contains the effective range and we use the effective range as a parameter for reproducing a reasonable neutron--matter EOS up to around the saturation density of symmetric matter. 
We have found that $r_s=$ -4.5 fm for the case  $|k_F \; a(k_F)|=1$ leads to an acceptable EOS (black dotted line in Fig. 4).

The symmetric matter  EOS will also be modified by the inclusion of the $t_1$ term in the interaction. 
We proceed as done previously: we readjust the parameters $t_0$, $t_3$, and $t_1$  to have a satisfactory EOS for symmetric matter and we keep these values unchanged. The adjusted parameters are $t_0$ = -1818 MeV fm$^3$, $t_3$ = 12970 MeV fm$^4$, $t_1$ = 15 MeV fm$^5$ and the corresponding EOS of symmetric matter is plotted in Fig. 1 within the model $t_0-t_1-t_3$.

\begin{figure}
\includegraphics[scale=0.35]{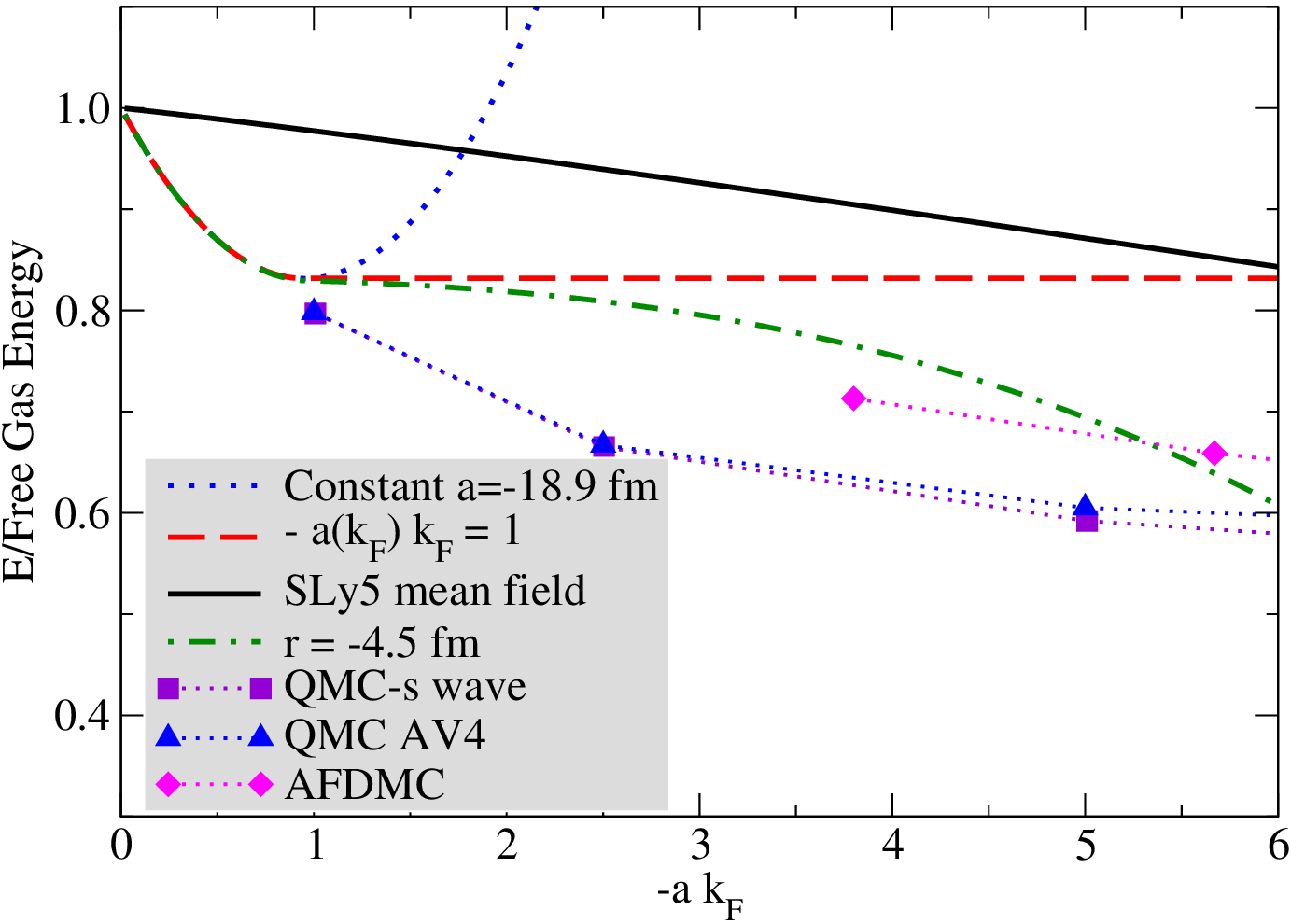}
\caption{Energy of neutron matter divided by the energy of a free Fermi gas as a function of $|a k_F|$, with $a=$ -18.9 fm. The black solid and the blue dotted curves represent the SLy5--mean--field and the LY (with only the first two terms and $a=$ -18.9 fm) EOSs, respectively. The red dashed and the green dot--dashed curves illustrate the EOSs obtained with a density--dependent scattering length with $|a(k_F) k_f|=1$, with only the first two terms and with the full expression of Eq. (\ref{ly}) ($r_s=$ -4.5 fm), respectively. 
The ab-initio Quantum Monte-Carlo (QMC)
with only s-wave and with the full AV4 interaction taken from Refs. \cite{Gez10,Car12} are shown respectively with purple squares and blue triangle. The AFDMC ab-initio results are taken from Ref. \cite{Gan08}.} 
\label{freeg}
\end{figure}

Let us now investigate in more detail the very low--density sector. Figure \ref{freeg} shows the energy of neutron matter divided by the free gas energy as a function of $-ak_F$ (with $a=-18.9$ fm).  
The curves corresponding to the first two terms of the Lee-Yang expansion with a constant scattering length (-18.9 fm) and with a density--dependent scattering length are shown together with the SLy5 EOS. In addition, the dot--dashed green curve illustrates the results obtained with the inclusion of the $k_F^5$--terms with $r_s=$-4.5 fm. We have already observed that this case gives a reasonable EOS at ordinary nuclear densities (Fig. 4).  We see now that, in addition, it well   
reproduces the correct low--density behavior. For instance, for $-a k_F \sim$ 6, the value of the energy divided by the free gas energy is $\sim$ 0.6, which is comparable to the microscopic QMC $s$--wave, the QMC AV4, or the AFDMC results reported in Ref. \cite{gandolfi,Gez10, Car12,Gan08}. 

Our exploratory study can be summarized as follows. We start with a functional based on the first two terms of the LY expansion, which is correct at very low densities but produces a wrong EOS for $\rho>10^{-6}$ fm$^{-3}$. Then, for higher densities, we depart from the matching to an effective--range expansion by imposing a density-dependent scattering length $a(k_F)$, which is constrained by a dimensionless quantity $C$ ($|k_F a(k_F)|=C\leq 1$). Empirical evidence shows that the effective range needs to enter as a free parameter in order to obtain a reasonable description of neutron matter up to the saturation density. The EOS of symmetric and neutron matter can be deduced by a simple functional which corresponds to a $t_0-t_1-t_3$ Skyrme interaction with density--dependent $x_i's$. Our functional may be employed into various applications for the description of neutron--rich systems at very low--density scales as well as of isospin--symmetric and asymmetric systems at ordinary nuclear density scales.

\begin{acknowledgments}
This project has received funding from
the European Unions Horizon 2020 research and innovation
program under grant agreement No. 654002.
\end{acknowledgments}

\end{document}